\documentclass[aps,pre,twocolumn,superscriptaddress,showpacs]{revtex4}
\usepackage{dcolumn}
\usepackage{graphicx}
\usepackage[dvips]{hyperref}
\usepackage[TS1,OT1,T1]{fontenc}

\begin{document}

 \newtheorem{thm}{Theorm}
 \newtheorem{defn}{Definition}
 \newtheorem{cor}{collory}
 \newtheorem{lem}{lemma}
 \newtheorem{prop}{proposition}

% Use the \preprint command to place your local institutional report
% number in the upper righthand corner of the title page in preprint mode.
% Multiple \preprint commands are allowed.
% Use the 'preprintnumbers' class option to override journal defaults
% to display numbers if necessary
%\preprint{}

%Title of paper
\title{Empirical analysis of the worldwide maritime transportation network}

% repeat the \author .. \affiliation  etc. as needed
% \email, \thanks, \homepage, \altaffiliation all apply to the current
% author. Explanatory text should go in the []'s, actual e-mail
% address or url should go in the {}'s for \email and \homepage.
% Please use the appropriate macro foreach each type of information

% \affiliation command applies to all authors since the last
% \affiliation command. The \affiliation command should follow the
% other information
% \affiliation can be followed by \email, \homepage, \thanks as well.

\author{Yihong Hu}
\email{051025007@fudan.edu.cn} \affiliation{School of management,
Fudan University, Shanghai 200433, China}
\author{Daoli Zhu}
\affiliation{School of management, Fudan University, Shanghai
200433, China} \affiliation{Shanghai logistics Institute, Shanghai
200433, China}
%Collaboration name if desired (requires use of superscriptaddress
%option in \documentclass). \noaffiliation is required (may also be
%used with the \author command).
%\collaboration can be followed by \email, \homepage, \thanks as well.
%\collaboration{}
%\noaffiliation

\date{\today}

\begin{abstract}
In this paper we present an empirical study of the worldwide
maritime transportation network (WMN) in which the nodes are ports
and links are container liners connecting the ports. Using the
different representations of network topology namely the space $L$
and $P$, we study the statistical properties of WMN including degree
distribution, degree correlations, weight distribution, strength
distribution, average shortest path length, line length distribution
and centrality measures. We find that WMN is a small-world network
with power law behavior. Important nodes are identified based on
different centrality measures. Through analyzing weighted cluster
coefficient and weighted average nearest neighbors degree, we reveal
the hierarchy structure and rich-club phenomenon in the network.
\end{abstract}

% insert suggested PACS numbers in braces on next line
% insert suggested keywords - APS authors don't need to do this
%\keywords{}

%\maketitle must follow title, authors, abstract, \pacs, and \keywords
\maketitle

% body of paper here - Use proper section commands
% References should be done using the \cite, \ref, and \label commands
\section{Introduction}
\label{introduction} The recent few years have witnessed a great
devotion to exploration and understanding of underlying mechanism of
complex systems as diverse as the Internet \cite{Barabasi}, social
networks \cite{Newman1} and biological networks \cite{Jeong}. As
critical infrastructure, transportation networks are widely studied.
Examples include airline \cite{Amaral,GuimeraPNAS,Wli, chiliping,
Guida}, ship \cite{Xinping1}, bus \cite{Sienkiewicz, Xinping2,
liping, Ferber}, subway \cite{Latorasubway} and railway \cite{Sen,
wli2} networks.

Maritime transportation plays an important role in the world
merchandize trade and economics development. Most of the large
volume cargo between countries like crude oil, iron ore, grain, and
lumber are carried by ocean vessels. According to the statistics
from United Nations \cite{maritimereport}, the international
seaborne trade continuously increased to 7.4 billion tons in 2006
with a robust annual growth rate of $4.3$ per cent. And over 70 per
cent of the value of world international seaborne trade is being
moved in containers.

Container liners have become the primary transportation mode in
maritime transport since 1950's. Liner shipping means the container
vessels travel along regular routes with fixed rates according to
regular schedules. At present most of the shipping companies adopt
hub-and-spoke operating structure which consists of hub ports,
lateral ports, main lines and branch lines, forming a complex
container transportation network system \cite{Rodrigue}.

Compared with other transportation networks, the maritime container
liner networks have some distinct features: (1) A great number of
the routes of container liners are circular. Container ships call at
a series of ports and return to the origin port without revisiting
each intermediate port. It's called pendulum service in container
transportation. While bus transport networks and railway networks
are at the opposite with most of buses or trains running
bidirectionally on routes. (2) The network is directed and
asymmetric due to circular routes. (3) Lines are divided into main
lines and branch lines. Main lines are long haul lines which
involves a set of sequential port calls across the oceans. Sometimes
long haul lines call at almost 30 ports. Branch lines are short haul
lines connecting several ports in one region to serve for main
lines.

We construct the worldwide maritime transportation network (WMN)
using two different network representations and analyze basic
topological properties. Our result shows that the degree
distribution follows a truncated power-law distribution in the space
$L$ and an exponential decay distribution in the space $P$. With
small average shortest path length 2.66 and high cluster coefficient
0.7 in the space $P$, we claim that WMN is a small world network. We
also check the weighted network and find the network has hierarchy
structure and "rich-club" phenomenon. Centrality measures are found
to have strong correlations with each other.

The rest of the paper is organized as follows: in Section
\ref{construction}, we introduce the database and set up the network
using two different network representations. In Section
\ref{basicproperty} various topological properties are studied
including degree distribution, degree correlations, shortest path
length, weight distribution and strength distribution etc. Section
\ref{structure} discloses the hierarchy structure by studying the
weighted and unweighted clustering and degree correlations.
Centrality measures correlations and central nodes' geographical
distribution are studied in Section \ref{centrality}. Section
\ref{conclusion} gives the conclusion.
% The Appendices part is started with the command \appendix;
% appendix sections are then done as normal sections
% \appendix

% \section{}
% \label{}
\section{Construction of the network}

\label{construction} We get the original data from a maritime
transport business database named CI-online \cite{CIonline} which
provides ports and fleet statistics of 434 ship companies in the
world. The data includes 878 sea ports and 1802 lines. The ports are
distributed in different regions and we list the number of ports in
each region in Table \ref{listofports}.
\renewcommand{\arraystretch}{1.2}
\begin{table}
\begin{center}
\caption{Number of sea ports by major geographic region
\label{listofports}}
\begin{tabular}{l c c c c}
\hline
Region & &  & No of sea ports & \\
\hline
Africa & & & 96 & \\
Asia and Middle East & & & 251 & \\
Europe & & & 311 &\\
North America & & & 61 &\\
Latin America & & & 96 &\\
Oceania & & & 63 &\\
Total & & & 878 &\\
\hline
\end{tabular}
\end{center}
\end{table}

\begin{figure}[th]
\begin{center}
\includegraphics[scale=0.35]{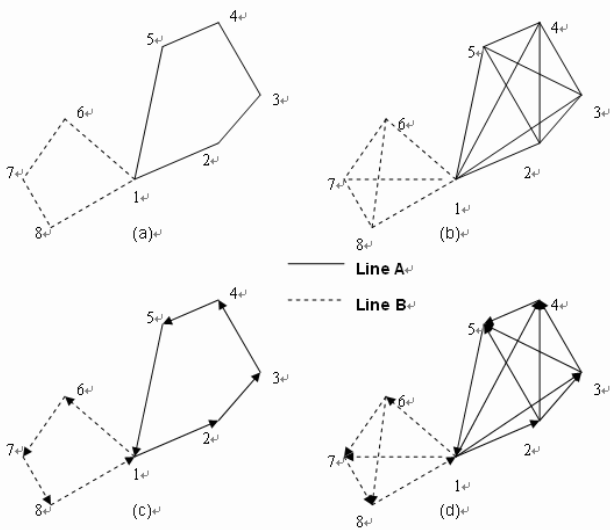}
\caption{Description of the space $L$ and the space $P$. (a) and (b)
are the undirected representations in the space $L$ and the space
$P$ respectively. (c) and (d) are the directed representations in
the space $L$ and the space $P$ respectively. In the space $L$, a
link is created between consecutive stops in one route. In the space
$P$ all ports that belong to the same route are connected. Line A
(solid line) and line B (dashed line) are two different pendulum
routes sharing one common node: the port No. 1. \label{spaceLandP}}
\end{center}
\end{figure}

To construct the worldwide maritime transportation network, we have
to introduce the concept of spaces $L$ and $P$ as presented in Fig.
\ref{spaceLandP}. The idea of spaces $L$ and $P$ is first proposed
in a general form in \cite{Sen} and later widely used in the study
of public bus transportation networks and railway networks. The
space $L$ consists of nodes being ports and links created between
consecutive stops in one route. Degree $k$ in the space $L$
represents the number of directions passengers or cargoes can travel
at a given port. The shortest path length in the space $L$ is the
number of stops one has to pass to travel between any two ports. In
the space $P$, two arbitrary ports are connected if there is a
container line traveling between both ports. Therefore degree $k$ in
the space $P$ is the number of nodes which can be reached without
changing the line. The shortest path length between any two nodes in
the space $P$ represents the transfer time plus one from one node to
another and thus is shorter than that in the space $L$.

Since WMN is a directed network, we extend the concept of spaces $L$
and $P$ to directed networks according to \cite{Xinping1}. See Fig.
\ref{spaceLandP}. Line A and B are two different pendulum routes
crossing at the port No. 1. (a) and (b) is the undirected network
representation. (c) and (d) is the respective directed version.

Based on the above concepts we establish the network under two
spaces represented by asymmetrical adjacent matrices $A^L$, $A^P$
and weight matrices $W^L$, $W^P$. The element $a_{ij}$ of the
adjacent matrix $A$ equals to 1 if there is a link from node $i$ to
$j$ or 0 otherwise. The element $w_{ij}$ of weight matrix $W$ is the
number of container lines traveling from port $i$ to port $j$.

We need to define the quantities used in this weighed and directed
network. We employ $k^{L}_{in}(i)$ and $k^{L}_{out}(i)$ to denote
in-degree, out-degree of node $i$ in the space $L$, and
$k^{L}_{un}(i)$ to represent undirected degree in the space $L$.
Similarly $k^{P}_{in}(i)$, $k^{P}_{out}(i)$ and $k^{P}_{un}(i)$ are
employed in the space $P$. Hence we have
\begin{equation}
k^{L}_{in}(i)=\sum_{j\neq i}a_{ji}^L
\end{equation}
\begin{equation}
k^{L}_{out}(i)=\sum_{j\neq i} a_{ij}^L
\end{equation}
\begin{equation}
k^{L}_{un}(i)=\sum_{j\neq i} (a_{ij}^L+a_{ji}^L)
\end{equation}
which also holds for the space $P$.

Strength is also divided into in-strength and out-strength. In the
space $L$ the in-strength of node $i$ is denoted by $S^{L}_{in}(i)$
and out-strength is denoted by $S^{L}_{out}(i)$. Undirected strength
(total strength) is represented by $S^{L}_{un}(i)$. They can be
calculated according to the following equations:
\begin{equation}
S^{L}_{in}(i)=\sum_{j\neq i}w_{ji}^L
\end{equation}
\begin{equation}
S^{L}_{out}(i)=\sum_{j\neq i} w_{ij}^L
\end{equation}
\begin{equation}
S^{L}_{un}(i)=\sum_{j\neq i} (w_{ij}^L+w_{ji}^L)
\end{equation}
which also holds for $S^{P}_{in}(i)$, $S^{P}_{out}(i)$,
$S^{P}_{un}(i)$ in the space $P$.

Other quantities like clustering coefficient and average nearest
neighbors degree also have different versions in directed and
weighted WMN. We employ $c^{L}_i$ and $c^{P}_i$ to denote the
unweighted clustering coefficient of node $i$ in the space $L$ and
$P$ respectively. Analogously $k_{nn,i}^{L}$ and $k_{nn,i}^{P}$ are
used to denote the average nearest neighbors degree of node $i$ in
the space $L$ and $P$ respectively. For weighted WMN we add
superscript $W$ to the above quantities and consequently they become
$c^{WL}_i$, $c^{WP}_i$, $k_{nn,i}^{WL}$ and $k_{nn,i}^{WP}$.

\begin{figure*}[th]
\begin{center}
\includegraphics[width=3.1in]{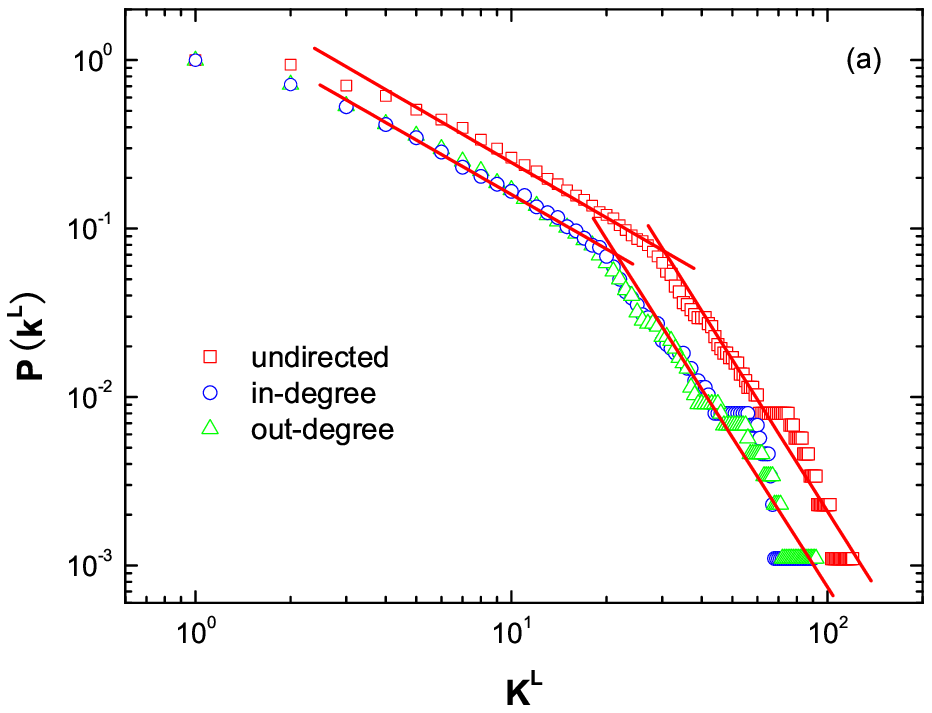}
\includegraphics[width=3.1in]{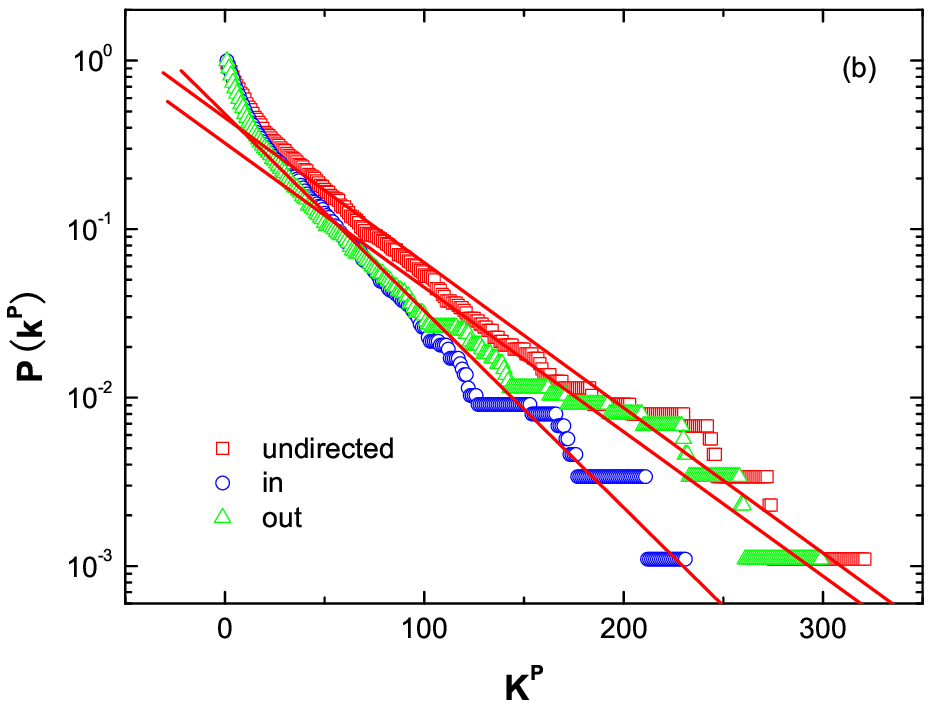}
\includegraphics[width=3.1in]{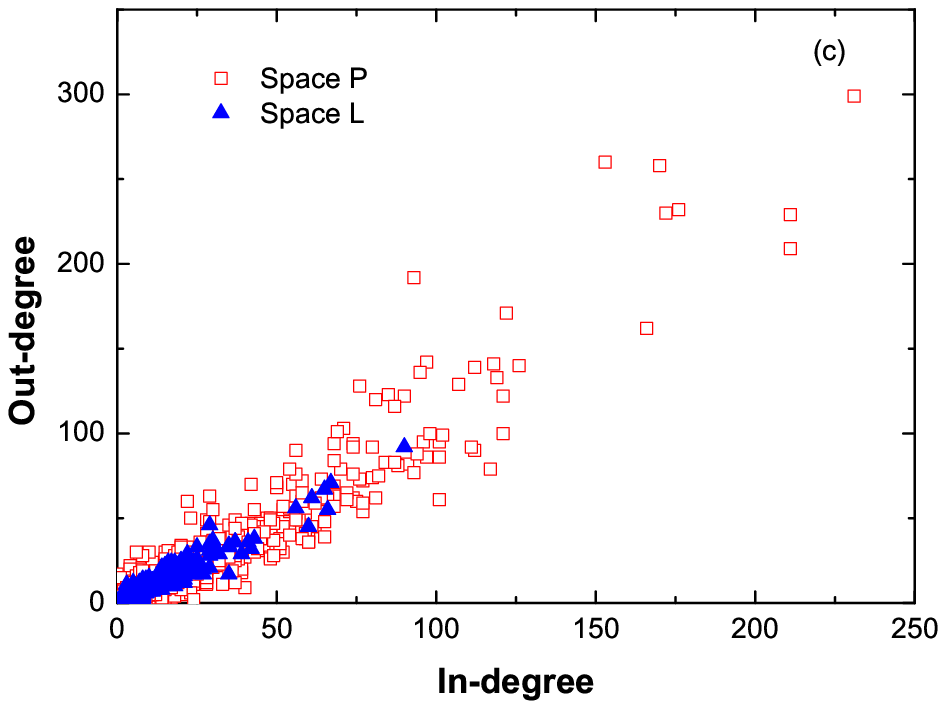}
\includegraphics[width=3.1in]{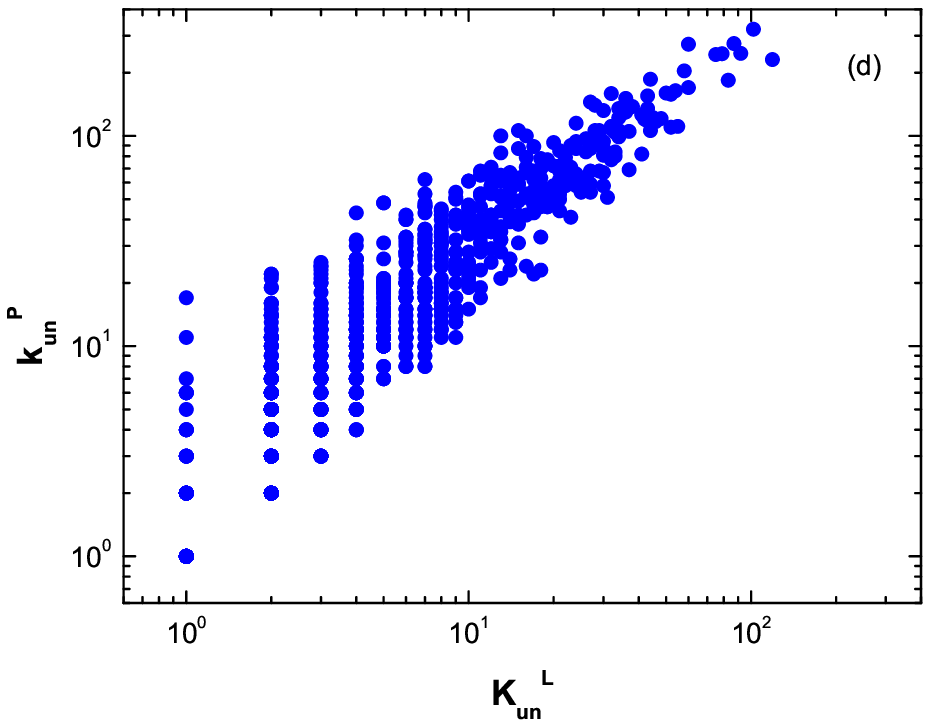}
\caption{(a) Degree distributions in the space $L$ obey truncated
power-law distributions with almost the same exponents. The turning
points are at $k=20$ and $k=30$ respectively. (b) Degree
distributions in the space $P$ all follow exponential distributions.
(c) Positive correlation between in-degree and out-degree. (d)
Unweighted degree under two spaces. Degree in the space $P$ is
larger than in the space $L$. \label{maritimedegree}}
\end{center}
\end{figure*}
\section{Topological properties}
\label{basicproperty}
\renewcommand{\arraystretch}{1.2}
\renewcommand{\tabcolsep}{9pt}
\begin{table}
\caption{Basic parameters for spaces $L$ and $P$. $n$ is the number
of nodes and $m$ is the number of links. $\langle k\rangle$ is the
average undirected degree. $\langle C\rangle$ is the average
unweighted cluster coefficient. $\langle l\rangle$ is the average
shortest path length.\label{maritimebasicprop}}
\begin{tabular}{l c c c c c }
\hline
Space & $n$ & $m$ & $\langle k_{un}\rangle$ & $\langle C\rangle$ & $\langle l\rangle$  \\
\hline
Space $L$ & 878 & 7955 & 9.04 & 0.4002 & 3.60 \\
Space $P$ & 878 & 24967 & 28.44 & 0.7061 & 2.66 \\
\hline
\end{tabular}
\end{table}

\subsection{degree distribution and degree correlations}
First we examine the degree distributions in two spaces. Fig.
\ref{maritimedegree} shows that in-degree, out-degree and undirected
degree distributions in the space $L$ all follow truncated power-law
distributions with nearly the same exponents. In-degree and
out-degree obey the function $P(k)\sim k^{-1.7}$ before $k=20$. When
$k>20$ their distribution curves bend down to the function $P(k)\sim
k^{-2.95}$. Unweighted degree in the space $L$ has the same
exponents of $-1.7$ and $-2.95$ but the critical point becomes
$k=30$. Truncated power-law degree distributions are often observed
in other transportation networks like the worldwide air
transportation network \cite{GuimeraPNAS}, China airport network
\cite{Wli}, U.S. airport network \cite{chiliping} and the Italian
airport network \cite{Guida}. It is explained in \cite{Amaral} that
the connection cost prevents adding new links to large degree nodes.
Analogous cost constraints also exist in the maritime transport
network. Congestion in hub ports often makes ships wait outside for
available berth for several days, which can cost ships extremely
high expense. Consequently new links are not encouraged to connect
to those busy ports.

While in the space $P$ three degree distributions all follow
exponential distributions $P(k)\sim e^{\alpha k}$. The parameters
are estimated to be $\alpha=0.0117$ for in-degree, and
$\alpha=0.0085$ for out-degree, $\alpha=0.0086$ for unweighted
degree. The property that degrees obey truncated power-law
distributions in the space $L$ and exponential distributions in the
space $P$ is identical to public transportation networks
\cite{Sienkiewicz,Xinping2} and railway networks \cite{Sen}.
Particularly the Indian railway network \cite{Sen} has exponential
degree distributions with the parameter $0.0085$ almost the same
with in-degree and out-degree distribution in WMN.

Next, the relation between in-degree and out-degree is studied. Fig.
\ref{maritimedegree}(c) is a plot of out-degree $k_{out}$ vs.
in-degree $k_{in}$. They have a positive correlation under two
spaces. Out-degree climbs when in-degree increases. Evidently the
in-out degree correlation is very strong.

Finally, we want to find out the relations between degrees under
different spaces. Fig. \ref{maritimedegree}(d) shows undirected
degree in the space $P$ $k_{un}^P$ vs. undirected degree in the
space $L$ $k_{un}^L$. All dots are above the diagonal, indicating
the undirected degree in the space $P$ is larger than in the space
$L$. It is understandable because of different definition of two
spaces topology. In the space $P$ all stops in the same route are
connected which surely increases degree of each node. Table
\ref{maritimebasicprop} lists the basic properties of WMN in two
spaces. Average undirected degree in the space $P$ is $28.44$, much
larger than average undirected degree 9.04 in the space $L$.

\subsection{Line length}
\begin{figure}[th]
\begin{center}
\includegraphics[scale=0.8]{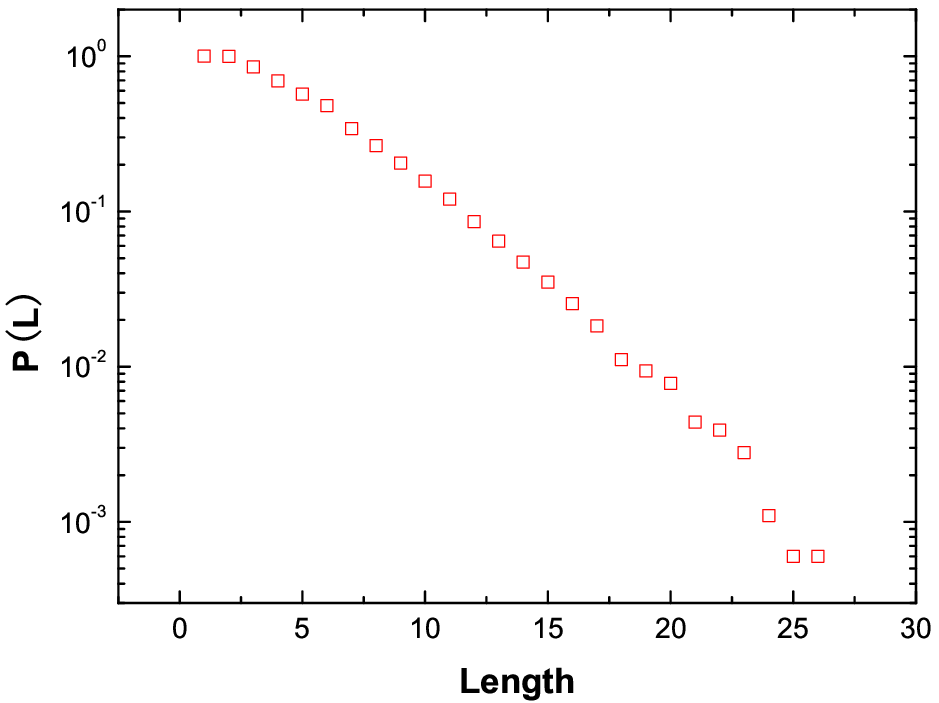}
\caption{Probability distribution of line length. It can be
approximated by a straight line in semi-log plot indicating an
exponential decay. \label{maritimelengthdist}}
\end{center}
\end{figure}

Let's denote line length, i. e. the number of stops in one line, as
$l$. In Fig. \ref{maritimelengthdist} the probability distribution
of line length $P(l)$ can be approximated as a straight line in the
semi-log picture representing an exponential decay distribution
$P(l)\sim e^{-\alpha l}$ with the parameter $\alpha=0.13$. It
indicates there are much more short haul lines than long haul lines
in maritime transportation. Long haul lines use large vessels and
travel long distance from one region to another region while short
haul lines as branch lines travel between several neighboring ports
and provide cargos to main lines. For example, the line consisting
of the following ports:
Shanghai-Busan-Osaka-Nagoya-Tokyo-Shimizu-Los
Angeles-Charleston-Norfolk-New York-Antwerp-Bremerhaven-
Thamesport-Rotterdam-Le Havre-New York-Norfolk-Charleston-Colon-Los
Angeles-Oakland-Tokyo-Osaka-Shanghai, is a typical long haul line
connecting main ports in Asia and Europe, calling at ports for 24
times.

\subsection{Shortest path length}
\begin{figure}[th]
\begin{center}
\includegraphics[scale=0.8]{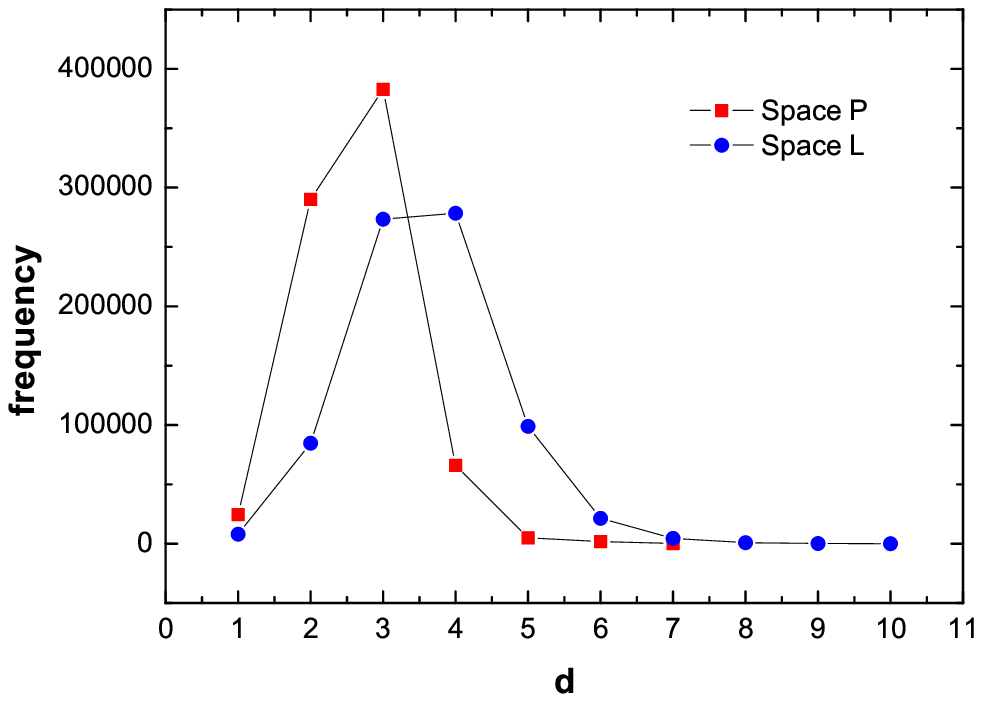}
\caption{Frequency distributions of shortest path length under two
spaces. The distribution in the space $L$ has a wider range than in
the space $P$.\label{maritimepathlength}}
\end{center}
\end{figure}

The frequency distributions of shortest path lengths $d$ in the
spaces $L$ and $P$ are plotted in Fig. \ref{maritimepathlength}. The
distribution in the space $L$ has a wider range than in the space
$P$. The average shortest path length is 3.6 in the space $L$ and
2.66 in the space $P$ (see Table \ref{maritimebasicprop}). This
means generally in the whole world the cargo need to transfer for no
more than 2 times to get to the destination. Compared with the
network size $N=878$, the shortest path length is relatively small.

\begin{figure*}[th]
\begin{center}
\includegraphics[width=3.1in]{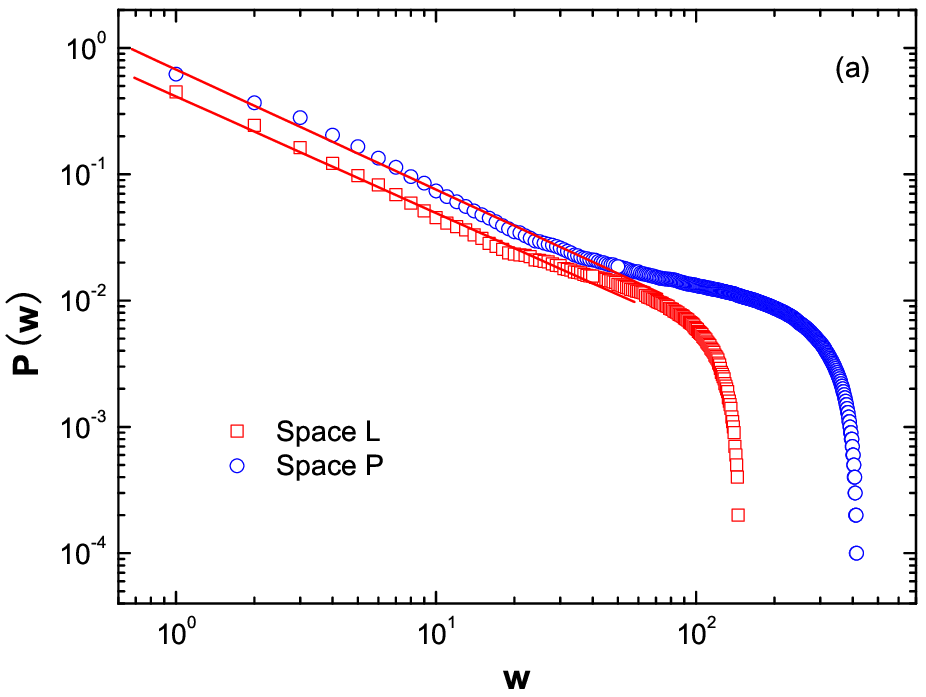}
\includegraphics[width=3.1in]{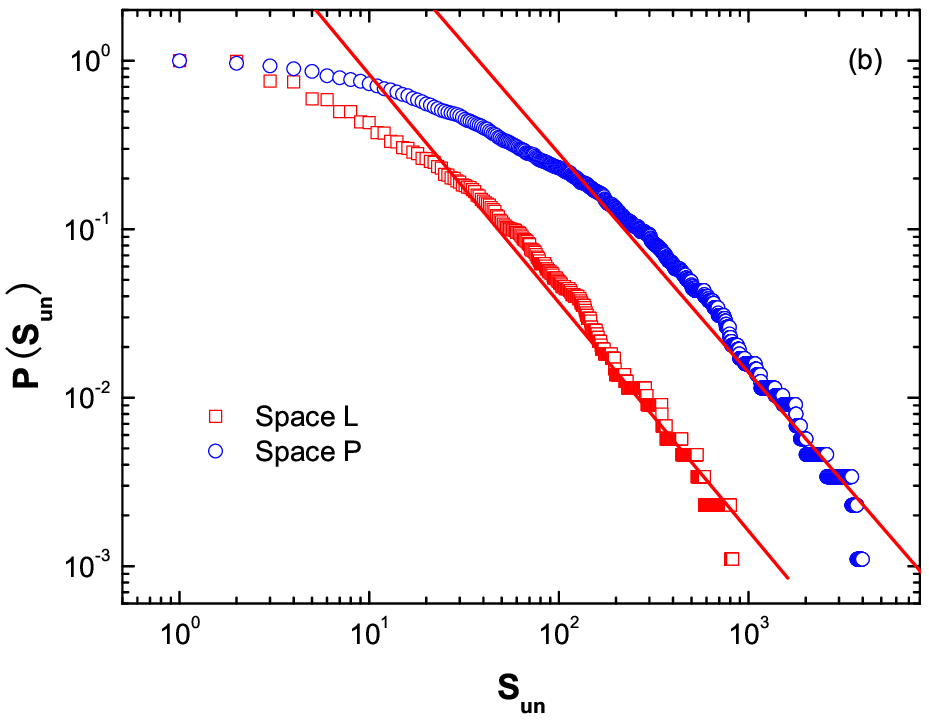}
\includegraphics[width=3.1in]{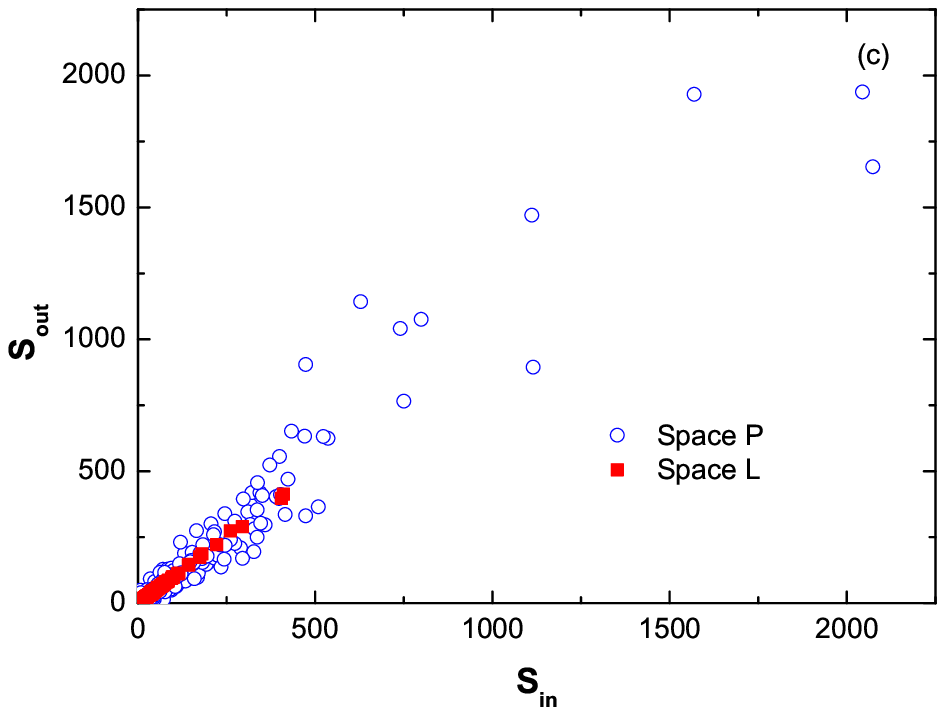}
\includegraphics[width=3.1in]{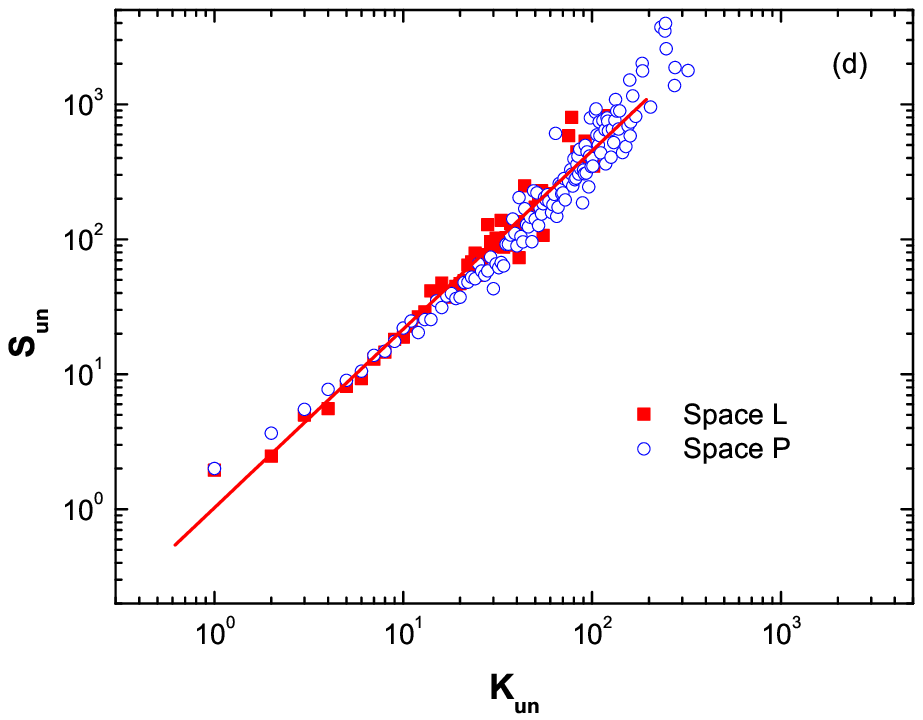}
\caption{(a) Weight distributions in two spaces. (b) Undirected
strength distributions in two spaces. (c) Relation between
in-strength and out-strength in two spaces. (d) Correlations between
strength and degree in two spaces. The slopes both equal to 1.3
approximately indicating nonlinear relationships in the spaces $L$
and $P$. \label{maritimeweightdist}}
\end{center}
\end{figure*}

\subsection{Weight and strength distribution}
Usually traffic on the transportation network is not equally
distributed. Some links have more traffic flow than others and
therefore play a more important role in the functioning of the whole
network. Weight should be addressed especially in transportation
networks. Here we study four properties of weighted WMN: weight
distribution, strength distribution, in-out strength relations and
the relations between strength and degree. The results are displayed
in Fig. \ref{maritimeweightdist}.

First we examine weight distributions. In Fig.
\ref{maritimeweightdist} (a) two weight distribution curves are
approximately straight declining lines before $w=40$. The power-law
distributions are estimated to be $P(w)\sim w^{-0.95}$ in the space
$P$ and $P(w)\sim w^{-0.92}$ in the space $L$.

Next, Fig. \ref{maritimeweightdist} (b) shows the undirected
strength distributions under two spaces both obey power-law behavior
with the same parameter. The functions are estimated to be $P(s)\sim
s^{-1.3}$.

And we also analyze the relation between in-strength and
out-strength in two spaces. As we can see from Fig.
\ref{maritimeweightdist} (c), in-out strength relation under the
space $L$ is plotted as almost a straight line while the relation
under the space $P$ has a slight departure from the linear behavior
at large $s_{in}$ values. But clearly they are positively
correlated.

Finally an important feature of weighted WMN, the relations between
strength and degree, is investigated. Under two spaces the relations
between undirected strength and undirected degree are both unlinear
with the slope of the line approximately $1.3$, which means the
strength increase quicker than the increase of degree. This is often
occured in the transportation networks and has its implication in
the reality. It's easy for the port with many container lines to
attract more lines to connect the port and thus to increase the
traffic more quickly.

\section{Hierarchy structure}
\label{structure}

In this section, we explore the network structure of WMN through
studying both the weighed and unweighted versions of cluster
coefficient and average nearest neighbors degree. Hierarchy
structure and "rich club" phenomenon are unveiled. We conjecture
this kind of structure is related to ship companies' optimal
behavior to minimize the transportation cost known as the
hub-and-spoke model in transportation industry.

\subsection{Clustering}
\begin{figure}[th]
\begin{center}
\includegraphics[width=3.1in]{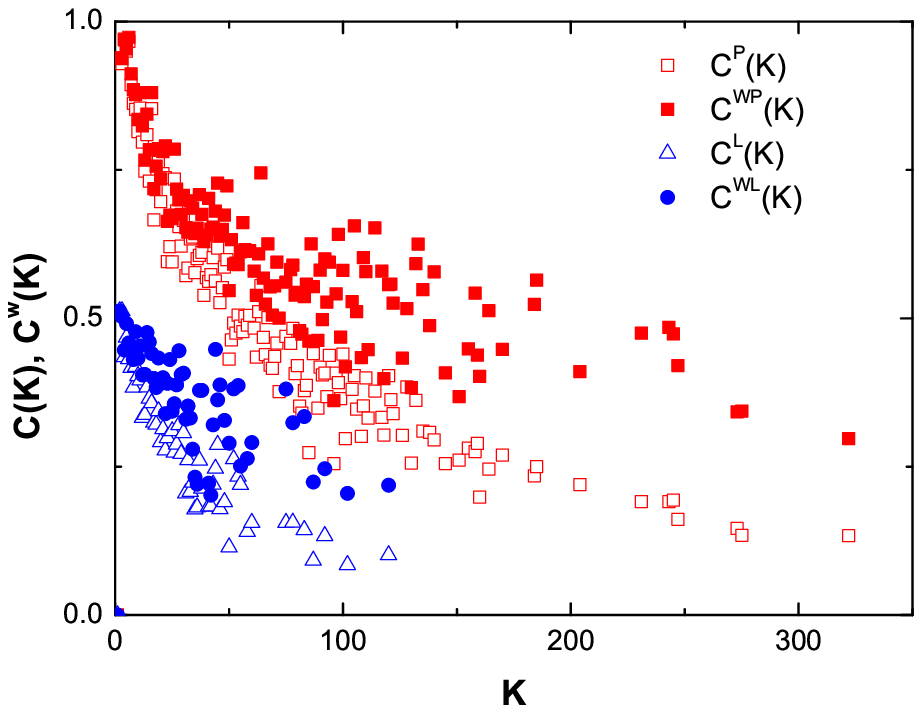}
\caption{Cluster coefficients. $C(k)$ under two spaces both exhibit
nontrivial behavior with decay curves as functions of degree $k$.
And weighted versions of cluster coefficient are larger than
unweighted cluster coefficient, indicating high traffic edges
between interconnected vertices.\label{maritimeck}}
\end{center}
\end{figure}

Cluster coefficient $c_i$ is used to measure local cohesiveness of
the network in the neighborhood of the vertex. It indicates to what
extent two individuals with a common friend are likely to know each
other. And $C(k)$ is defined as cluster coefficient averaged over
all vertices with degree $k$.

We plot $C(k)$ in Fig. \ref{maritimeck}. Either in the space $L$ or
in the space $P$, $C(k)$ exhibits a highly nontrivial behavior with
a decay curve as a function of degree $k$, signaling a hierarchy
structure in which low degrees belong generally to well
interconnected communities (high clustering coefficient), while hubs
connect many vertices that are not directly connected (small
clustering coefficient).

$C^P(k)$ lies above $C^L(k)$ and the average cluster coefficient of
the network in the space $P$ is 0.7 larger than  0.4 in the space
$L$. This can be explained by the fact that in the space $P$ each
route gives rise to a fully connected subgraph. With high cluster
coefficient 0.7 and small average shortest path length 2.66 in the
space $P$, we conclude that the WMN, as expected, has the
small-world property.

Weighed quantities for clustering and assortativity measures are
first proposed in \cite{Barrat3}. Through the case study of WAN and
SCN, \cite{Barrat3} demonstrates that the inclusion of weight and
their correlations can provide deeper understanding of the
hierarchical organization of complex networks. The weighted cluster
coefficient is defined as:
\begin{equation}
c^w_i=\frac{1}{s_i(k_i-1)}\sum_{j,h}\frac{w_{ij}+w_{ih}}{2}a_{ij}a_{ih}a_{jh}
\end{equation}
which takes into account the importance of the traffic or
interaction intensity on the local triplets. And we define $C^w(k)$
as the weighted cluster coefficient averaged over all vertices with
degree $k$. In real weighted network we may have two opposite cases
of the relation between $C^w(k)$ and $C(k)$. If $C^w(k)>C(k)$ in the
network, interconnected triplets are more likely formed by the edges
with larger weights. If $C^w(k)<C(k)$ the largest interactions or
traffic is occurring on edges not belonging to interconnected
triplets.

In Fig. \ref{maritimeck} we report weighted cluster coefficient
under two spaces. Evidently the weighted cluster coefficient
$C^{wP}(k)$ and $C^{wL}(k)$ are both above the corresponding
unweighted cluster coefficient, i. e. $C^{wP}(k)\geq C^P(k)$, and
$C^{wL}(k)\geq C^L(k)$. This indicates some closely interconnected
nodes with large degrees have the edges with larger weights among
themselves. In other words, high-degree ports have the tendency to
form interconnected groups with high-traffic's links, thus balancing
the reduced clustering. This is so called "rich-club" phenomenon.

\subsection{Assortativity}
\begin{figure}[th]
\begin{center}
\includegraphics[width=3.1in]{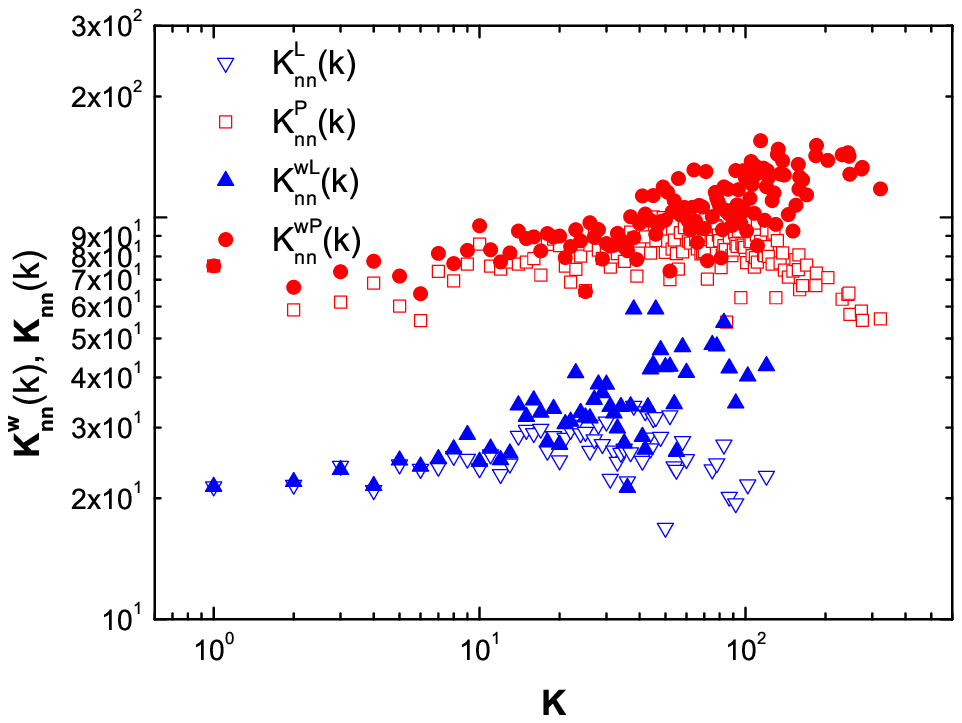}
\caption{Average degree of the nearest neighbors as functions of
$k$. Inclusion of weight changes the behavior of $k_{nn}(k)$:
assortative behavior in the small degree range but disassortative
behavior in large degree range, to definite assortative behavior of
$k^w_{nn}(k)$ in the whole $k$ spectrum. \label{maritimeknni}}
\end{center}
\end{figure}

There is another important quantity to probe the networks'
architecture: the average degree of nearest neighbors, $k_{nn}(k)$,
for vertices of degree $k$. Average nearest neighbors degree of a
node $i$ is defined as:
\begin{equation}
k_{nn,i}=\frac{1}{k_i}\sum_{j}a_{ij}k_j
\end{equation}
Using $k_{nn,i}$, one can calculate the average degree of the
nearest neighbors of nodes with degree $k$, denoted as $k_{nn}(k)$.
The networks are called assortative if $k_{nn}(k)$ is an increasing
function of $k$, whereas they are referred to as disassortative when
$k_{nn}(k)$ is a decreasing function of $k$. As suggested in
\cite{Barrat3}, weighted version of average degree of nearest
neighbors is calculated by:
\begin{equation}
k^w_{nn,i}=\frac{1}{s_i}\sum_{j}a_{ij}w_{ij}k_{j}
\end{equation}
From this definition we can infer that $k^w_{nn,i}>k_{nn,i}$ if the
links with the larger weights are pointing to the neighbors with
larger degrees and $k^w_{nn,i}<k_{nn,i}$ in the opposite case.

Both the weighted and unweighted average degree of nearest neighbors
under two spaces are plotted in Fig. \ref{maritimeknni}. The curve
of $k^P_{nn}(k)$ lies above the curve of $k^L_{nn}(k)$. The
$k^L_{nn}(k)$ and $k^P_{nn}(k)$ grow with the increase of degrees at
small degrees but decline when degrees are large. The unweighted
network exhibits assortative behavior in the small degree range but
disassortative behavior in large range.

When we turn to the $k^{wL}_{nn}(k)$ and $k^{wP}_{nn}(k)$, the
weighted analysis provides us a different picture. We can see that
under two spaces the weighted average degree of nearest neighbors
exhibits a pronounced assortative behavior in the whole $k$
spectrum. Since the number of WMN nodes is 878, this conforms with
the empirical finding in \cite{Sienkiewicz} that public transport
networks are assortative when the number of nodes in the network
$N>500$ and disassortative when $N<500$.

From the above discussion we can see that the inclusion of weight
changes the behavior of cluster coefficient and average degree of
nearest neighbors. This property is identical to the worldwide
airport network \cite{Barrat3} and North America airport network
\cite{Barrat4}. In both the airline transportation and maritime
transportation networks, high traffic is associated to hubs and
high-degree ports (airports) tend to form cliques with other large
ports (airports). Their similar organization structure may have a
similar underlying mechanisms. We conjecture that this is related to
the hub-and-spoke structure which is widely adopted in practice by
airline companies or ship companies to achieve the objective of
minimizing the total transportation cost \cite{OKelly, Robinson,
Moura,Bryan}.

\begin{figure}[th]
\begin{center}
\includegraphics[scale=0.4]{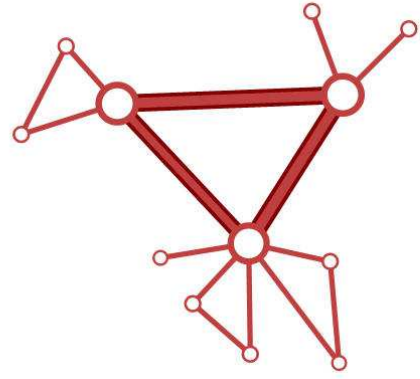}
\caption{A classical hub-and-spoke structure in maritime
transportation. There are three central vertices having very strong
links (high traffic) with each other and several nodes having weak
links (low traffic) with hubs. It has the same property of
$C^w(k)>C(k),k^w_{nn}(k)>k_{nn}(k)$ with that of WMN.
\label{maritimehubandspoke}}
\end{center}
\end{figure}

Fig. \ref{maritimehubandspoke} describes a typical hub-and-spoke
structure which consists of three interconnected hubs and other
nodes allocated to a single hub. In maritime transportation main
liners travel between hubs handling large traffic while branch
liners visit the hub's neighboring ports to provide cargo for the
main lines. This structure allows the carriers to consolidate the
cargo in larger vessels to lower the transportation cost. This
simple structure has the similar property of
$C^w(k)>C(k),k^w_{nn}(k)>k_{nn}(k)$ with that of WMN. And it also
displays rich-club phenomenon. We think it worths investigating the
relations between ship companies' optimal behavior and the real
transportation network's hierarchy structure and rich-club property.

\section{Centrality measures}

\begin{figure}[th]
\begin{center}
\includegraphics[width=3.1in]{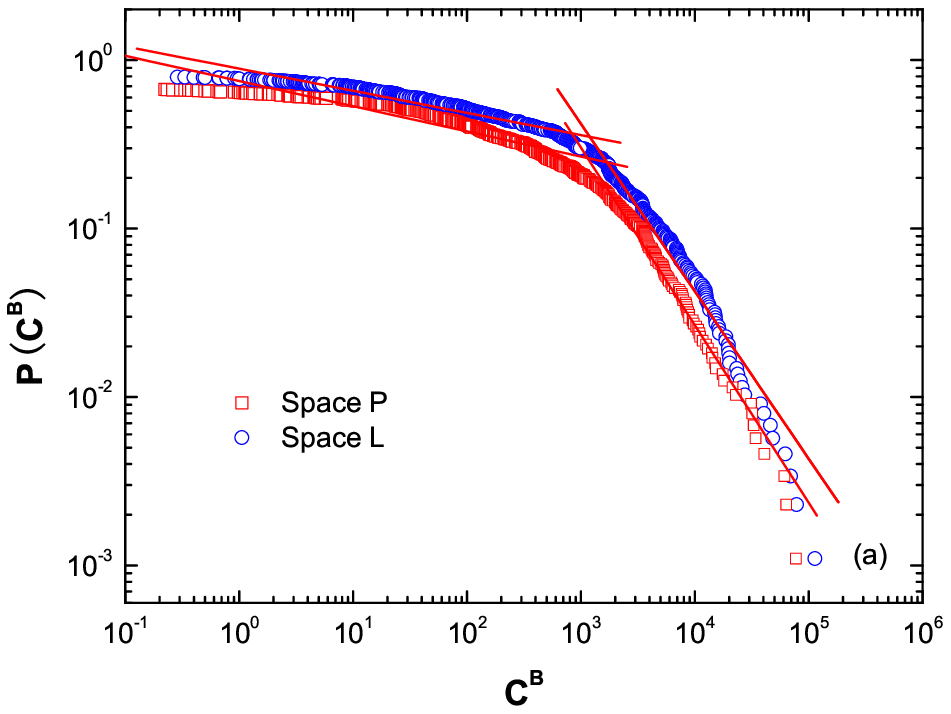}
\includegraphics[width=3.1in]{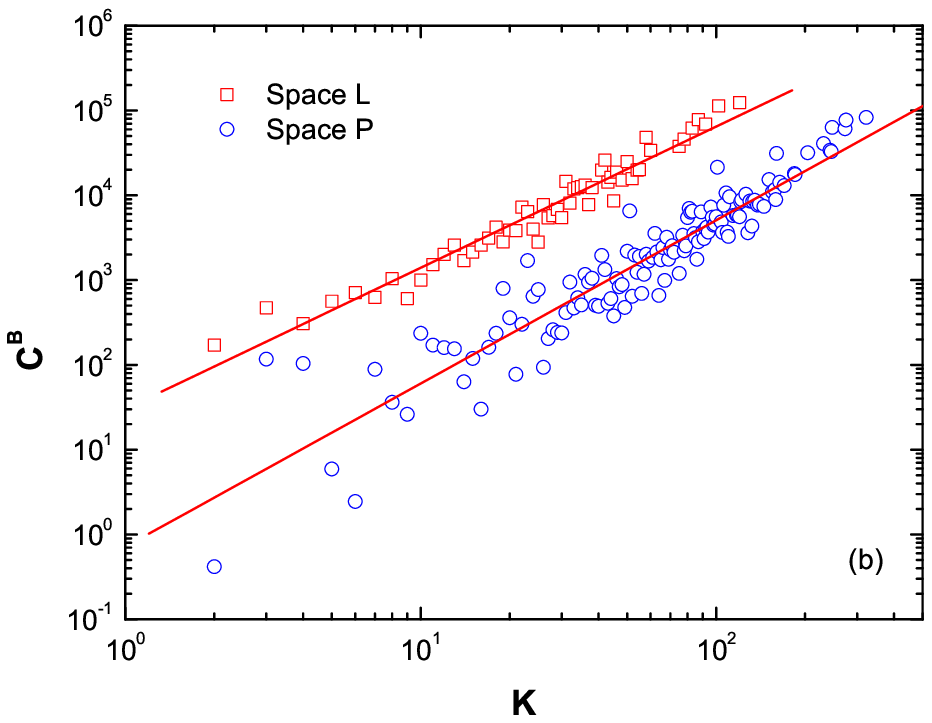}
\caption{(a) Probability distribution of betweenness obeys truncated
power-law distribution. (b) Betweenness is a straight line in
log-log picture indicating a power-law relation with degree. The
exponent in the space $L$ is estimated to be 1.66 and that in the
space $P$ is estimated to be 1.93. \label{maritimecentrality}}
\end{center}
\end{figure}

In this section we analyze two centrality measures in social network
analysis \cite{Wasserman}: degree and betweenness. The most
intuitive topological measure of centrality is given by the degree:
more connected nodes are more important. The distribution and
correlations of degree has been discussed in Section IV.

Betweenness centrality is defined as the proportion of the shortest
paths between every pair of vertices that pass through the given
vertex $v$ towards all the shortest paths. It is based on the idea
that a vertex is central if it lies between many other vertices, in
the sense that it is traversed by many of the shortest paths
connection couples of vertices. Hence we have
\begin{equation}
C_i^B=\frac{1}{(N-1)(N-2)}\sum_{j,k} \frac{n_{jk}(i)}{n_{jk}}
\end{equation}
where $n_{jk}$ is the number of shortest paths between $j$ and $k$,
and $n_{jk}(i)$ is the number of shortest paths between $j$ and $k$
that contain node $i$.

Correlations between two centrality measures are presented in Fig.
\ref{maritimecentrality}(b). In both the spaces there is a clear
tendency to a power-law relation with degree $k$: $C^B(k)\sim
k^\alpha$ with $\alpha=1.66$ in the space $L$ and $\alpha=1.93$ in
the space $P$. The power-law correlations between degree and
betweenness is also found in bus transportaiton networks
\cite{Sienkiewicz, Ferber} and ship transport network
\cite{Xinping1}. It's worth noting that this power-law relations
together with the truncated scale-free behavior of the degree
distribution implies that betweenness distribution should follow a
truncated power law. This behavior is clearly identified in Fig.
\ref{maritimecentrality}(a). We find the betweenness centrality has
two-regime power-law behavior $P(C^B)\sim C^{-\alpha}$. For the two
spaces, exponents are almost the same: $\alpha=0.14$ at small degree
regime and $\alpha=1.0$ at large degree regime.

The power-law relations between degree and betweenness suggest that
they are consistent with each other. It is proved in the comparison
of each port's degree and betweenness. The 25 most connected ports
are listed in Table. \ref{20ports}. Singapore is the most busy ports
in the world with the largest degree and betweenness. Antwerp and
Bushan are the second and third either in degree or in betweenness
measures. Only 5 ports in these ports are not listed in the 25 most
central ports in betweenness measure. WMN is not like the case of
the worldwide airline network \cite{GuimeraPNAS} which has anomalous
centrality due to its multicommunity structure. The difference may
due to the fact that there are less geographical and political
constraints in maritime transportation than in air transportation.
Ships can travel longer distance than airplanes. And airports are
usually classified into international and domestic airports and
international airlines are limited to connect international airports
instead of domestic airports. So there are distinct geographically
constrained communities in WAN. In the maritime transportation there
are no such constraints. Sea ports basically are all international
ports with the possibility to connect to any other sea ports in the
world.

\label{centrality}
\renewcommand{\tabcolsep}{2pt}
\begin{table}
\begin{center}
\caption{The 25 most connected ports in the worldwide maritime
transportation network.
* These ports are not among the 25 most central ports. \label{20ports}}
\begin{tabular}{c l c c c}
\hline
Rank & Ports & Degree & Betweenness & Region\\
\hline
1 & Singpore      & 120 & 124110.1258 & Asia\\
2 & Antwerp    & 102 & 113368.6161 & Europe \\
3 & Bushan        & 92& 69094.7490 & Asia\\
4 & Rotterdam      & 87 & 78097.8754 &Europe\\
5 & Port Klang      & 83 & 62111.6226 & Asia\\
6 & Hongkong        & 78 & 46072.9799 & Asia\\
7 & Shanghai        & 75 & 37748.4316 & Asia\\
8 & Hamburg      & 60 & 40362.4625 & Europe\\
9 & Valencia  & 60 & 27346.0956 & Europe\\\
10 & Le Havre & 58 & 48231.1636 & Europe\\\
11 & Gioia Tauro & 55 & 20148.4667 & Europe\\
12 & Yokohama     & 54& 19716.7287 &Asia\\
13 & Kaohsiung     & 52& 15363.5781 & Asia\\
14 & Port Said*     & 52& 16121.8476 & Africa\\
15 & Bremerhaven & 50& 24998.9088 & Europe\\
16 & Colombo*   & 48 & 15128.362 & Asia\\
17 & Tanjung Pelepas &46 & 18735.1267 & Asia\\
18 & Jeddah*     & 45& 8570.7724 & Middle East\\
19 & Jebel Ali & 44& 20137.5255 & Middle East\\
20 & Ningbo*       & 44& 12315.9515 & Asia\\
21 & Algeciras & 43& 18701.2619 & Europe\\
22 & Barcelona*    & 43& 15256.3903 & Europe\\
23 & Kobe       & 43& 8869.7313 & Asia\\
24 & New York       & 42& 25935.6692 & North America\\
25 & Kingston       & 41& 23281.3223 & Latin America\\
\hline
\end{tabular}
\end{center}
\end{table}

\begin{figure*}[th]
\begin{center}
\includegraphics[scale=0.4]{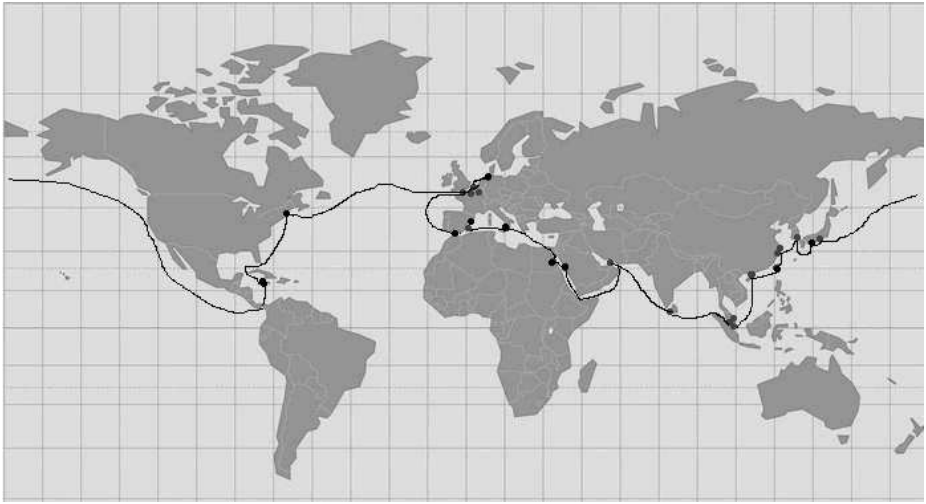}
\caption{The geographical distribution of 25 most connected ports.
They are located along east-west lines, including 13 ports in Asia
and Middle East, 9 in Europe, 1 in Africa, 1 in North America and 1
in Latin America. \label{maritimeworldmap}}
\end{center}
\end{figure*}

In Fig. \ref{maritimeworldmap} we plot the 25 most connected ports
on the world map. They show unbalanced geographical distribution
mainly located in Asia and Europe, including 13 ports in Asia and
Middle East, 1 in Africa, 9 in Europe, 1 in North America and 1 in
Latin America. Particularly they are located along the east-west
lines. Lines in maritime transportation are usually divided into
east-west lines, north-south lines and south-south lines
\cite{Rodrigue,Fremont}. The fact that 25 most connected ports in
the world are in east-west trade routes represents rapid growth and
large trade volume in Europe-America, Asia-America and Asia-Europe
trade \cite{maritimereport}.

\section{Conclusion}
\label{conclusion} In this paper we have presented an empirical
study of the worldwide maritime transportation network (WMN) under
different representations of network topology. We study the
statistical properties of WMN and find that WMN is a small world
network with power law behavior. There are strong correlations in
degree-degree, strength-degree and betweenness-degree relations.
Central nodes are identified based on different centrality measures.
Based on the analysis of weighted cluster coefficient and weighted
average nearest neighbors degree, we find that WMN has the same
hierarchy structure and "rich-club" phenomenon with WAN. We
conjecture that this structure is related to optimal behavior both
existing in air transportation and maritime transportation. So our
future research direction is the evolution modeling of WMN using
optimal behavior to reproduce real properties in WMN.

\section*{Acknowledgment}
The work was supported by Natural Science Foundation of China and
USA ffgg (NSFC 70432001).
% Specify following sections are appendices. Use \appendix* if there
% only one appendix.
%\appendix
%\section{}

% If you have acknowledgments, this puts in the proper section head.
%\begin{acknowledgments}
% put your acknowledgments here.
%\end{acknowledgments}

\appendix

\end{document}